%% file: main.tex
\documentclass[a4paper,twocolumn]{article}
\pdfoutput=1

\ifdefined\XeTeXversion
        \usepackage{eulervm}
        \usepackage[cm-default]{fontspec}%
        \usepackage{xunicode}
        \usepackage{xltxtra}
\fi

\usepackage{abstract}
\usepackage{authblk}
\usepackage{ccaption}
\usepackage[width=18cm]{geometry}
\usepackage{amsmath}
\usepackage{booktabs}
\usepackage{color}
\usepackage{graphicx}
\usepackage{listings}
\usepackage[caption=false]{subfig}
\usepackage{url}
\usepackage[pdfborder={0 0 0},colorlinks=true,citecolor=red,linkcolor=blue]{hyperref}

\captiondelim{. }
\captionnamefont{\sc\bfseries}
\captiontitlefont{\itshape}

\definecolor{light-gray}{gray}{0.95}
\title{A Global Optimisation Toolbox for Massively Parallel Engineering Optimisation}
\author{Francesco Biscani\footnote{\href{mailto::bluescarni@gmail.com}{bluescarni@gmail.com}}}
\author{Dario Izzo\footnote{\href{mailto:dario.izzo@esa.int}{dario.izzo@esa.int}}}
\author{Chit Hong Yam\footnote{\href{mailto:chit.hong.yam@esa.int}{chit.hong.yam@esa.int}}}
\affil{European Space Agency -- Advanced Concepts Team \\ European Space Research and Technology Centre (ESTEC)}

\begin{document}

\maketitle

\begin{abstract}
A software platform for global optimisation, called PaGMO, has been developed within the 
Advanced Concepts Team (ACT) at the European Space Agency, and was recently released as an open-source project. 

PaGMO is built to tackle high-dimensional global optimisation problems, and it has been successfully used to find solutions to real-life engineering problems among which the preliminary design of interplanetary
spacecraft trajectories - both chemical (including multiple flybys and deep-space maneuvers) and low-thrust (limited, at the 
moment, to single phase trajectories), the inverse design of nano-structured radiators and the design of non-reactive controllers for planetary rovers.

Featuring an arsenal of global and local optimisation algorithms (including genetic 
algorithms, differential evolution, simulated annealing, particle swarm optimisation, compass search, improved harmony search, and various interfaces to libraries for local optimisation
such as SNOPT, IPOPT, GSL and NLopt), PaGMO is at its core a C++ library which employs an object-oriented architecture providing a clean and easily-extensible
optimisation framework. Adoption of multi-threaded programming ensures the efficient exploitation of modern 
multi-core architectures and allows for a straightforward implementation of the island model paradigm, in which multiple 
populations of candidate solutions asynchronously exchange information in order to speed-up and improve the optimisation 
process. In addition to the C++ interface, PaGMO's capabilities are exposed to the high-level language Python, so that it is 
possible to easily use PaGMO in an interactive session and take advantage of the numerous scientific Python libraries 
available.
\end{abstract}

\input{introduction}
\input{pagmo}

\input{results}

\input{conclusions.tex}

\bibliographystyle{abbrv}
\bibliography{biblio}

\end{document}

%% file: introduction.tex
\section{Introduction}

With the introduction of mass-produced multi-core architectures, personal computers are becoming increasingly capable of performing parallel computations. Yet, the effort to parallelize algorithms is time consuming and often not attractive, especially in scientific computing where software reuse is not as spread a practice as in other fields of computing. The open-source project PaGMO, (Parallel Global Multiobjective Optimiser), aims at filling this gap for optimisation algorithms providing, through a generalization of the so called island model (i.e. a coarse grained approach to parallelization of genetic algorithms) to all types of algorithms (population based and not), a simple experimenting platform that allows scientists to easily code algorithms and problems without having to care at all about the underlying parallelization that is provided `for free' by the PaGMO infrastructure . The resulting software platform, participating to the Google initiative Summer of Code 2010, is described in this paper together with application examples to real life engineering problems of interest to aerospace engineers.

Recent results in global optimisation algorithms applied to the design of chemically-propelled interplanetary 
trajectories have shown how a straightforward application of off-the-shelf optimisation algorithms does not suffice to find satisfactory 
solutions for the most complex cases such as the Messenger trajectory or the Cassini or the TandEM trajectory \cite{izzo}. While a wise 
use of the algorithms still provides useful information also in these most complex cases, the final optimal solutions need a 
substantial amount of engineering knowledge to be found. In this paper we show how the use of PaGMO allows different 
algorithms to cooperate to the solution of the same interplanetary trajectory problem, allowing to find solutions also in the most difficult 
cases in a reasonable time. In particular, we demonstrate a fully automated search of the solution space based on the use of Differential 
Evolution, Simulated Annealing and local search in a cooperative fashion. Information is exchanged asynchronously between 
the solvers operating in parallel CPUs via the implementation of a generalized migration operator offered by PaGMO. We test this search strategy in the 
case of the Cassini, TandEM and Messenger trajectories as defined in the European Space Agency Global Trajectory 
optimisation Problems database (GTOP) \cite{gtop1,gtop2}. We show that the 
algorithms are able to locate interesting regions of the search space and in particular to find the possible resonances. In the case of TandEM
and Cassini, an automatic pruning strategy is able to successfully identify the best known solutions. In the case of Messenger, the automated search locates a large number of possible solution clusters (due to the possible resonances at Mercury). A second run of the search focussed on a particular one of these clusters holds satisfactory results and  is, in particular, able to find the same strategy adopted by the actual Messenger mission. A last example is then presented, where 4406 simpler interplanetary trajectories are optimised (each five times) taking dvantage of PaGMO's parallelization capabilities in a reasonably short time to locate preliminarly good targets for an asteroid sample return mission in the 2020-2050 time frame.

%% file: pagmo.tex
\section{PaGMO}

PaGMO is an optimisation framework developed within the Advanced Concepts Team of the European Space Agency.
Written in C++, PaGMO aims to provide an extensible infrastructure for defining optimisation problems (nonlinear, continuous, integer, mixed-integer,
box-constrained, nonlinearly constrained, multi-objective optimisation is supported),
coupled with a wide arsenal of global and local optimisation algorithms - some of them coded directly within PaGMO, others called
from external libraries through thin wrappers. At the time of this writing, PaGMO provides the following optimisation algorithms:
\begin{itemize}
 \item global and local optimisation algorithms coded directly within PaGMO, including a simple genetic algorithm \cite{sga}, differential evolution \cite{de},
 particle swarm optimisation \cite{pso},
 adaptive neighbourhood simulated annealing \cite{sa_corana},
 improved harmony search \cite{ihs},
 compass search \cite{cs},
 monotonic basin hopping \cite{mbh},
 generalised multistart and
 Monte Carlo search \cite{monte-carlo};
 \item wrapper for SNOPT \cite{snopt};
 \item wrapper for IPOPT \cite{ipopt};
 \item wrappers for algorithms from the NLopt library \cite{nlopt}, including Subplex \cite{subplex} (an extension of the classical Nelder-Mead method),
  COBYLA \cite{cobyla}
  and BOBYQA \cite{bobyqa};
 \item wrappers for algorithms from the GSL library \cite{gsl}, including
  the Broyden-Fletcher-Goldfarb-Shanno (BFGS) method \cite{bfgs},
  Fletcher-Reeves and Polak-Ribi\`{e}re nonlinear conjugate gradient methods \cite{conj_grad} and
  the classical Nelder-Mead method \cite{nelder_mead};
 \item wrappers for algorithms from the SciPy library \cite{scipy} (only available in the Python bindings), including
 fmin (Nelder-Mead), L-BFGS-B \cite{lbfgsb},
 sequential least-square programming \cite{slsqp}
  and truncated Newton method \cite{tnc}.
\end{itemize}

\begin{figure*}[ht]
 \begin{center}
  \includegraphics[width=14cm]{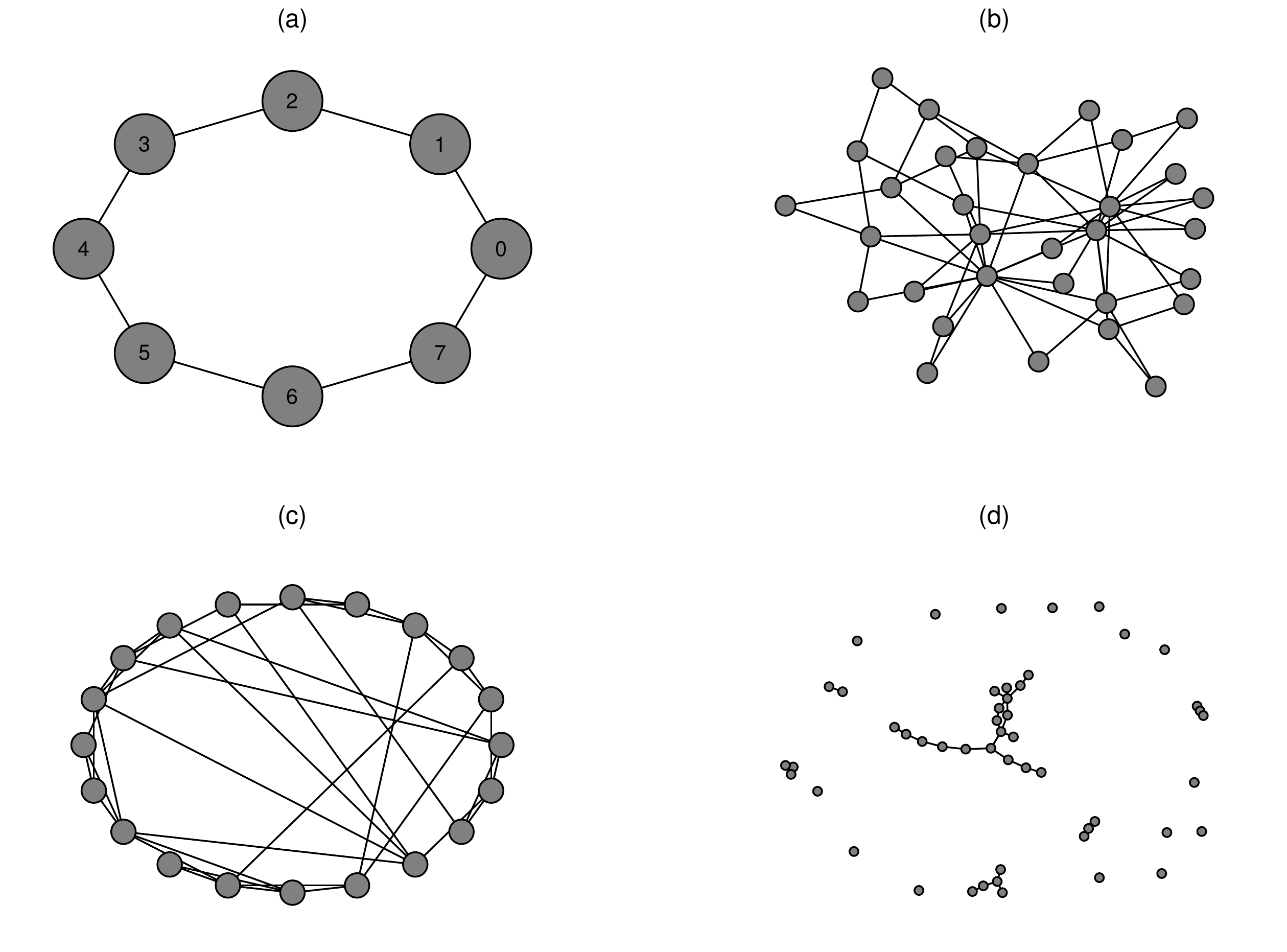}
  \caption{A selection of topologies available in PaGMO: ring topology (a), Barab\'{a}si-Albert model (b), Watts-Strogatz model (c) and Erd\H{o}s-R\'{e}nyi $G\left(n,p \right)$ random graph (d).}
  \label{fig:topologies}
 \end{center}
\end{figure*}

PaGMO provides automatic parallelisation of the optimisation process via a coarse-grained approach based on the island model \cite{island_model}, in which multiple optimisation
instances of the same problem are launched at the same time, asynchronously exchanging information and improving the overall convergence properties of the optimisation. In PaGMO's implementation of
the island model, each optimisation instance (i.e., each island) is launched in a separate thread of execution, thus automatically taking advantage of modern multiprocessor machines. The connections
between islands are respresented by a graph topology in which each node corresponds to an island and the edges represent routes through which candidate solutions can be communicated from one island to
the other. The graph topologies can be either constructed by manually adding nodes and edges, or they can be selected among those already coded within PaGMO, including:
\begin{itemize}
 \item popular topologies in the context of parallel population-based optimisation, such as fully connected, torus, cartwheel, lattice, hypercube,
 broadcast, and various types of ring topologies;
 \item small-world network topologies, such as the Barab\'{a}si-Albert \cite{ba_model} and Watts-Strogatz \cite{ws_model} models;
 \item $G\left(n,p \right)$ Erd\H{o}s-R\'{e}nyi random graph \cite{er_model};
 \item custom topologies (such as the wheel rim topology described in \S\ref{sec:examples}).
\end{itemize}
Some of the topologies available in PaGMO are visualised in Figure \ref{fig:topologies}. Full control over the fine-grained details of the 
migration strategy (e.g., migration frequency and rate, selection and replacement policies) is provided. A preliminary study of the impact 
the topology on the optimisation process can be found in \cite{topology}.
The class that contains the set of islands collaborating in an optimisation process, the topology and the migration policies is known in PaGMO as an \emph{archipelago}.

PaGMO ships with a number of implemented optimisation problems readily available for use, such as:
\begin{itemize}
 \item classical continuous test functions, such as Rastrigin, Rosenbrock \cite{rosenbrock}, Schwefel, Griewank, Branin, Himmelblau, Lennard-Jones potential \cite{lj_potential} and Levy5;
 \item constrained continuous test functions from \cite{luksan_vlcek};
 \item integer programming problems: Golomb ruler \cite{golomb}, 0-1 knapsack problem \cite{knapsack};
 \item multi-objective optimisation test problems from \cite{nsga-ii};
 \item all the chemical interplanetary spacecraft trajectory problems from the European Space Agency's GTOP database \cite{gtop1,gtop2};
 \item an interplanetary multiple gravity assist low-thrust problem.
\end{itemize}

PaGMO's C++ capabilities are exposed to the high-level language Python, so that it is possible to instantiate problems, algorithms, topologies and islands from either a script or an interactive Python session. It is also possible to
define new problems and algorithms directly from Python, thus allowing on one hand to rapidly prototype and evaluate new ideas, and on the other  to leverage the rich ecosystem of freely-available scientific Python modules
(e.g., numerical integrators, machine learning libraries, computer algebra systems, etc.). Coupled with the matplotlib plotting module and the enhanced Python shell IPython, PaGMO's Python bindings (which have been called PyGMO)
offer a user-friendly interactive graphical experience.

%% file: results.tex
\section{Some examples}
\label{sec:examples}

As an example of the use of PaGMO to solve engineering problems we report here the results
of the application of the optimisation strategy described in the previous section to four trajectory optimisation selected problems. The first three problems are taken from the European Space Agency Global Trajectory optimisation (GTOP) database \cite{gtop1,gtop2}. The problems selected are among the most difficult proposed in the database and are
 included in the basic PaGMO distribution. They all are
 box-constrained, continuous, single objective optimisation problems, representing a multiple gravity
 assist interplanetary trajectory with one deep space maneuver allowed in each trajectory leg. The search space includes launch windows spanning decades (see the GTOP database for the precise definitions of the allowed bounds on the launch and fly-by dates). The fourth problem is a simpler problem admitting though a large number of different instances. We take advantage of PaGMO parallelization to find solutions to 4406 different instances of the problem in a reasonable computing time.
 
\paragraph{Experimental setup}

\begin{figure}
 \begin{center}
  \includegraphics[width=7.7cm]{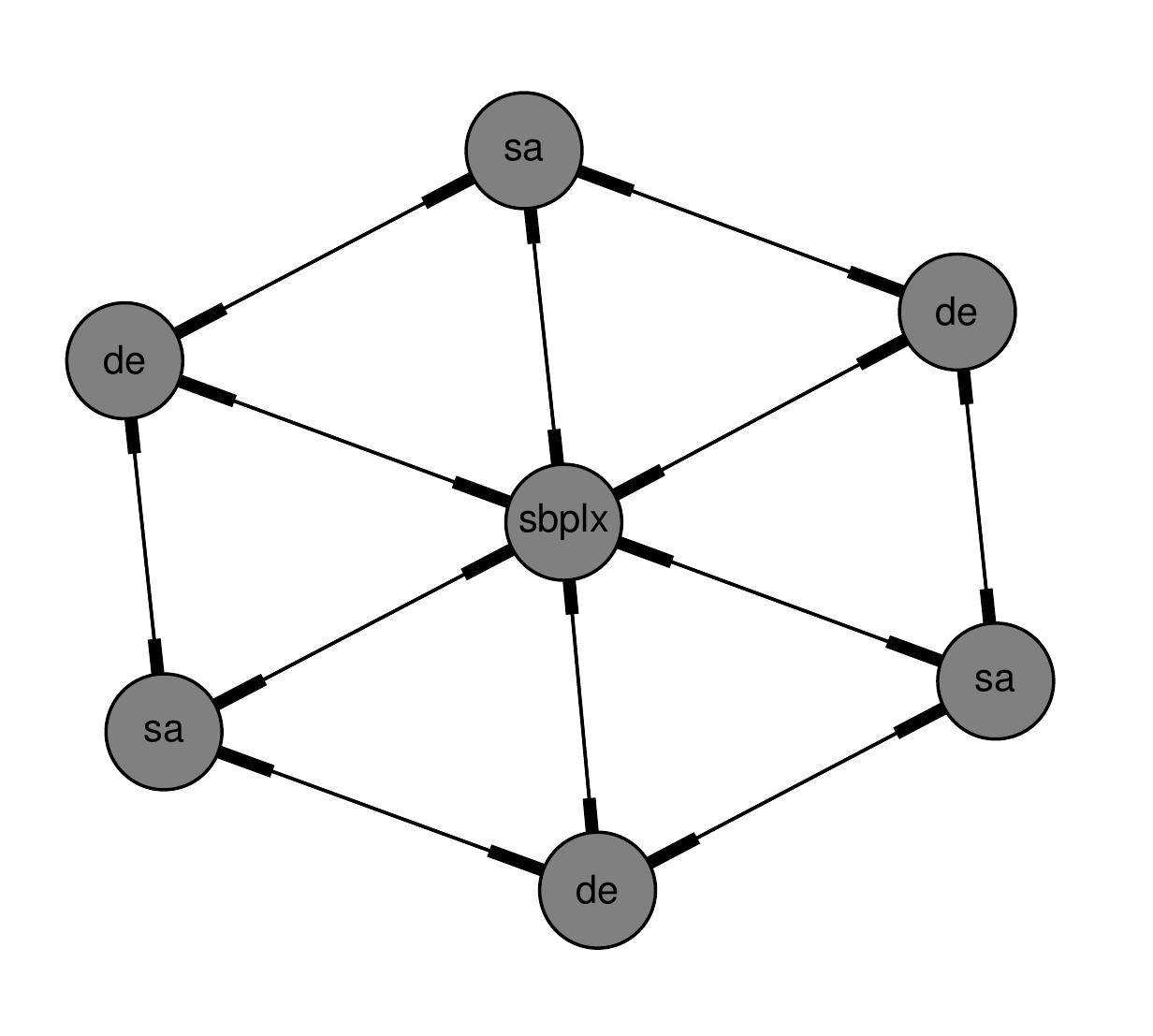}
  \caption{Experimental setup: an archipelago with wheel rim topology, global optimisation algorithms on the outer ring (simulated annealing and differential evolution)
  and a local optimisation algorithm in the inner island (Subplex).}
  \label{fig:rim}
 \end{center}
\end{figure}

All the optimisation problems were set up in an archipelago of 5-7 islands (depending on the number of available cores on the machine at the time of the experiment),
equipped with a wheel rim topology (see Figure \ref{fig:rim}). The wheel rim topology consists of a classical bidirectional ring topology with an additional island at the
center, fully connected to all the other islands. We chose to deploy global optimisation algorithms (namely, adaptive neighbourhood simulated annealing from \cite{sa_corana} and
differential evolution from \cite{de}) on the ring, and a local optimisation algorithm (namely, the Subplex algorithm from \cite{subplex} as implemented in 
\cite{nlopt}) in the center.

The motivations behind these choices are the following:
\begin{itemize}
 \item the ring topology is a proven and popular choice in the context of parallel population-based algorithms, as shown for instance in
 \cite{Homayounfaretal03, Starkweatheretal91, GordonWhitley93, CantuPaz00, CantuPazMejiaOlvera94, Izzoetal09, galapagos};
 \item the additional island in the center receives through migration the best results of the global optimisation algorithms in the ring,
 refines them through a local search, and migrates them back to the ring. Its role is hence, on one hand, to improve the results of the global search,
 and on the other to inject back diversified candidate solutions into the ring;
 \item regarding the choice of the algorithms, both simulated annealing and differential evolution have proven to be effective for the optimisation of
 interplanetary spacecraft trajectories (as shown for instance in \cite{Izzoetal09}), whereas the derivative-free Subplex method, a refinement of the classical Nelder-Mead
 algorithm, is particularly suited for the noisy and multi-modal objective functions appearing in these optimisation problems.
\end{itemize}

In order to give an example of use of PaGMO, we reproduce here the Python code necessary to perform one optimisation run with the setup described above:
\begin{lstlisting}
# Import the PyGMO classes
from PyGMO import * (*@\label{code:import}@*)

# Instantiate the algorithms
sa = algorithm.sa_corana(10000,1,0.01) (*@\label{code:algo_start}@*)
de = algorithm.de(500,0.8,0.9)
local = algorithm.nlopt_sbplx(500,1e-4) (*@\label{code:algo_end}@*)

# Instantiate the problem
prob = problem.messenger_full() (*@\label{code:prob}@*)

# Build the archipelago
a = archipelago(topology.rim()) (*@\label{code:archi_start}@*)
a.push_back(island(prob,local,1,1.0,migration.worst_r_policy())) (*@\label{code:archi_local}@*)
a.push_back(island(prob,sa,1)) (*@\label{code:algo_ring_start}@*)
a.push_back(island(prob,de,20))
a.push_back(island(prob,sa,1))
a.push_back(island(prob,de,20))
a.push_back(island(prob,sa,1))
a.push_back(island(prob,de,20)) (*@\label{code:archi_end}@*) (*@\label{code:algo_ring_end}@*)

# Perform evolution twenty times
a.evolve(20) (*@\label{code:archi_evolve}@*)
a.join() (*@\label{code:archi_join}@*)
\end{lstlisting}
Detailed explanation:
\begin{itemize}
 \item on line \ref{code:import}, all the PaGMO classes are imported into the current namespace;
 \item on lines \ref{code:algo_start}-\ref{code:algo_end}, the algorithms are instantiated;
 \item on line \ref{code:prob}, the problem (in this case the full Messenger problem) is instantiated;
 \item on lines \ref{code:archi_start}-\ref{code:archi_end}, the archipelago is instantiated:
 \begin{itemize}
  \item on line \ref{code:archi_start}, an empty archipelago with rim topology is created;
  \item on line \ref{code:archi_local}, the central island is created and inserted in the archipelago
  with the \lstinline!push_back()! method. The island is constructed from the problem \lstinline!prob! and the algorithm \lstinline!local!, it contains one single individual, has a probability of
  accepting the migrating individuals of 100\% and replacement policy \lstinline!migration.worst_r_policy()!, which will unconditionally replace the worst individual in the island with the incoming
  individuals. This island needs a non-default replacement policy because we want it to optimise \emph{every} candidate solution coming from the ring, whereas the default behaviour would be to accept
  migrating individuals only if they would improve upon the worst individuals present in the population (which is the behaviour frequently desired for population-based algorithms);
  \item on lines \ref{code:algo_ring_start}-\ref{code:algo_ring_end}, the ring islands are created and inserted into the archipelago. The simulated annealing islands operate on populations of a single
  individual, whereas the differential evolution islands are instantiated with a population of 20 individuals. The default migration policies are in these cases appropriate;
 \end{itemize}
 \item on line \ref{code:archi_evolve}, the optimisation process is started by calling the \lstinline!evolve()! method of the archipelago. The argument passed to the \lstinline!evolve()! method, in this case 20,
 means that each algorithm on each island is called 20 times with the parameters passed in the constructors on lines \ref{code:algo_start}-\ref{code:algo_end}. E.g., in case of differential evolution, it means that
 the algorithm is run for $500 \cdot 20 = 10000$ generations, with weight coefficient equal to 0.8 and crossover probability equal to 0.9. Migration is allowed to happen at the end of each one of the 20 internal iterations of the
 \lstinline!evolve()! method;
 \item on line \ref{code:archi_join}, the archipelago is joined, meaning that the flow of the program stops until the optimisation run started on line \ref{code:archi_evolve} has concluded. Since PaGMO runs asynchronously
 each algorithm in a separate thread, the \lstinline!evolve()! call on line \ref{code:archi_evolve} will return almost immediately -- the optimisation process having forked in the background. The \lstinline!join()! call
 blocks the program until the optimisation has finished.
\end{itemize}

\paragraph{Optimisation strategy}

For the first three problems we adopted the following optimisation strategy:
\begin{enumerate}
 \item we instantiated an archipelago with rim topology as described above and let the optimisation run for a fixed amount of time;
 \item at the end of each optimisation run, we recorded the best candidate solution produced and then reset the archipelago with randomly-chosen decision vectors.
\end{enumerate}
Step 1. and 2. where repeated multiple times, thus producing a collection of
optimised candidate solutions. The cluster pruning algorithm described in \cite{izzo} was then run on the collection of candidate solutions, returning
new problem bounds in which top decision vectors are contained.
The new bounds
are then used to launch other rounds of multistart optimisations.

For the sample return problem, which involves the solution of different instances of a simpler problem, a single run of the optimisation algorithms in the archipelago is good enough and thus no
cluster pruning was used.

\begin{table*}[ht]
\centering
\begin{minipage}[t]{5.35cm}
\subfloat[][]{
\centering
\begin{tabular}{ll}
        \noalign{\smallskip}\toprule\noalign{\smallskip}
        \multicolumn{2} {c}{Departure}  \\
        \noalign{\smallskip}\midrule\noalign{\smallskip}
        Epoch &   12/11/1997 \\
        $V_{\infty}$ & 3.254 km/s \\
        \noalign{\smallskip}\midrule\noalign{\smallskip}
        \multicolumn{2} {c}{Cruise}  \\
        \noalign{\smallskip}\midrule\noalign{\smallskip}
                DSM $\Delta V$ & 484 m/s\\
        Venus fly-by & 29/04/1998 \\
        DSM $\Delta V$ & 399 m/s\\
                Venus fly-by & 27/06/1999 \\
        Earth fly-by & 19/08/1999  \\
                Jupiter fly-by & 31/03/2001  \\
        
           \noalign{\smallskip}\midrule\noalign{\smallskip}
        \multicolumn{2} {c}{Arrival}  \\
        \noalign{\smallskip}\midrule\noalign{\smallskip}
        Epoch & 05/2007\\
           $V_{\infty}$ & 4.25 km/s \\
           \noalign{\smallskip}\midrule\noalign{\smallskip}
           Total flight time &  9.4 years\\
\noalign{\smallskip}\bottomrule
\end{tabular}
\label{tab:cassini2}
}
\end{minipage}
\begin{minipage}[t]{5.35cm}
\subfloat[][]{
\centering
\begin{tabular}{ll}
        \noalign{\smallskip}\toprule\noalign{\smallskip}
        \multicolumn{2} {c}{Departure}  \\
        \noalign{\smallskip}\midrule\noalign{\smallskip}
        Epoch & 15/11/2021 \\
        $V_{\infty}$ & 3.34 km/s \\
        Declination & 3.1 deg \\
        \noalign{\smallskip}\midrule\noalign{\smallskip}
        \multicolumn{2} {c}{Cruise}  \\
        \noalign{\smallskip}\midrule\noalign{\smallskip}
        Venus fly-by & 30/04/2022 \\
        Earth fly-by & 04/04/2023  \\
           DSM $\Delta V$ & 167 m/s\\ 
           Earth fly-by & 25/06/2026\\
           \noalign{\smallskip}\midrule\noalign{\smallskip}
        \multicolumn{2} {c}{Arrival}  \\
        \noalign{\smallskip}\midrule\noalign{\smallskip}
        Epoch & 03/07/2031\\
           $V_{SOI}$ & 0.676 km/s \\
           \noalign{\smallskip}\midrule\noalign{\smallskip}
           Total flight time &  9.63 years\\
        \midrule
\noalign{\smallskip}
        \multicolumn{2} {c}{Spacecraft}  \\
        \noalign{\smallskip}\midrule\noalign{\smallskip}
        Daparture mass & 2085.44 kg \\
        Arrival mass & 1476.03 kg \\
        $I_{sp}$        &       312 s\\
        \noalign{\smallskip}\bottomrule\noalign{\smallskip}
\end{tabular}
\label{tab:tandem}
}
\end{minipage}
\begin{minipage}[t]{5.35cm}
\subfloat[][]{
\centering
\begin{tabular}{ll}
        \noalign{\smallskip}\toprule\noalign{\smallskip}
        \multicolumn{2} {c}{Departure}  \\
        \noalign{\smallskip}\midrule\noalign{\smallskip}
        Epoch & 14/08/2005 \\
        $V_{\infty}$ & 3.95 km/s \\
        \noalign{\smallskip}\midrule\noalign{\smallskip}
        \multicolumn{2} {c}{Cruise}  \\
        \noalign{\smallskip}\midrule\noalign{\smallskip}
        Venus fly-by & 20/10/2006 \\
           DSM $\Delta V$ & 343 m/s\\ 
           Venus fly-by & 04/06/2007 \\
           DSM $\Delta V$ & 567 m/s\\ 
           Mercury fly-by & 12/01/2008\\
            DSM $\Delta V$ & 91 m/s\\ 
                   Mercury fly-by & 04/10/2008\\
            DSM $\Delta V$ & 224 m/s\\ 
                   Mercury fly-by & 28/09/2009\\
            DSM $\Delta V$ & 179 m/s\\ 
           \noalign{\smallskip}\midrule\noalign{\smallskip}
        \multicolumn{2} {c}{Arrival}  \\
        \noalign{\smallskip}\midrule\noalign{\smallskip}
        Epoch & 03/2011\\
           $V_{MOI}$ & 0.905 km/s \\
           \noalign{\smallskip}\midrule\noalign{\smallskip}
           Total flight time &  5.59 years\\
\noalign{\smallskip}\bottomrule
\end{tabular}
\label{tab:messenger}
}
\end{minipage}
\caption{Details of the best trajectories found for the first three interplanetary trajectory problems: problem::cassini\_2 \subref{tab:cassini2}, problem::tandem(6,10) \subref{tab:tandem} and
problem::messenger\_full \subref{tab:messenger}.}
\end{table*}

 \subsection{Results on problem::cassini\_2}
 
 \begin{figure*}
 \centering
  \subfloat[][]{\includegraphics[width=8cm]{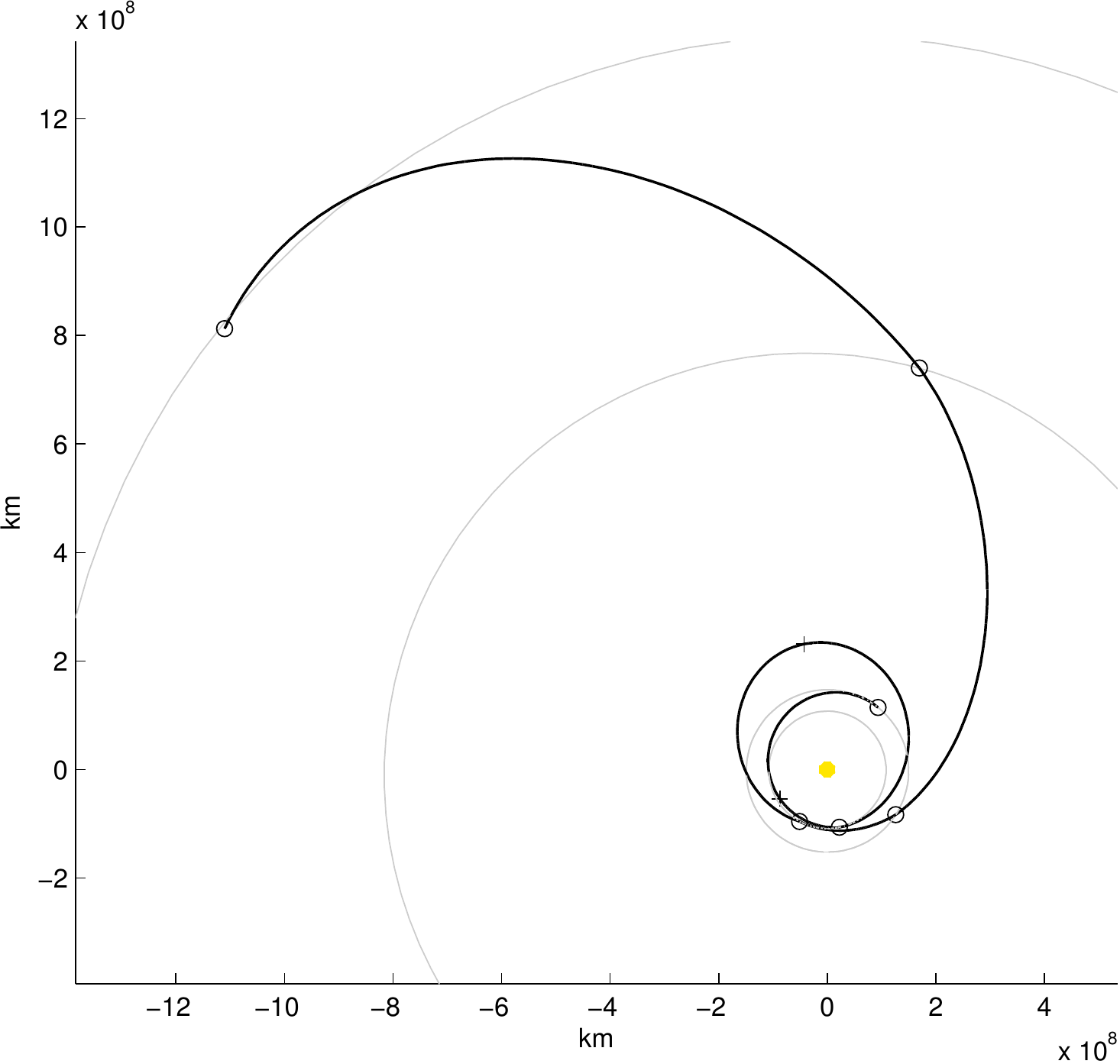}\label{fig:cassini2}}
  \qquad
  \subfloat[][]{\includegraphics[width=8cm]{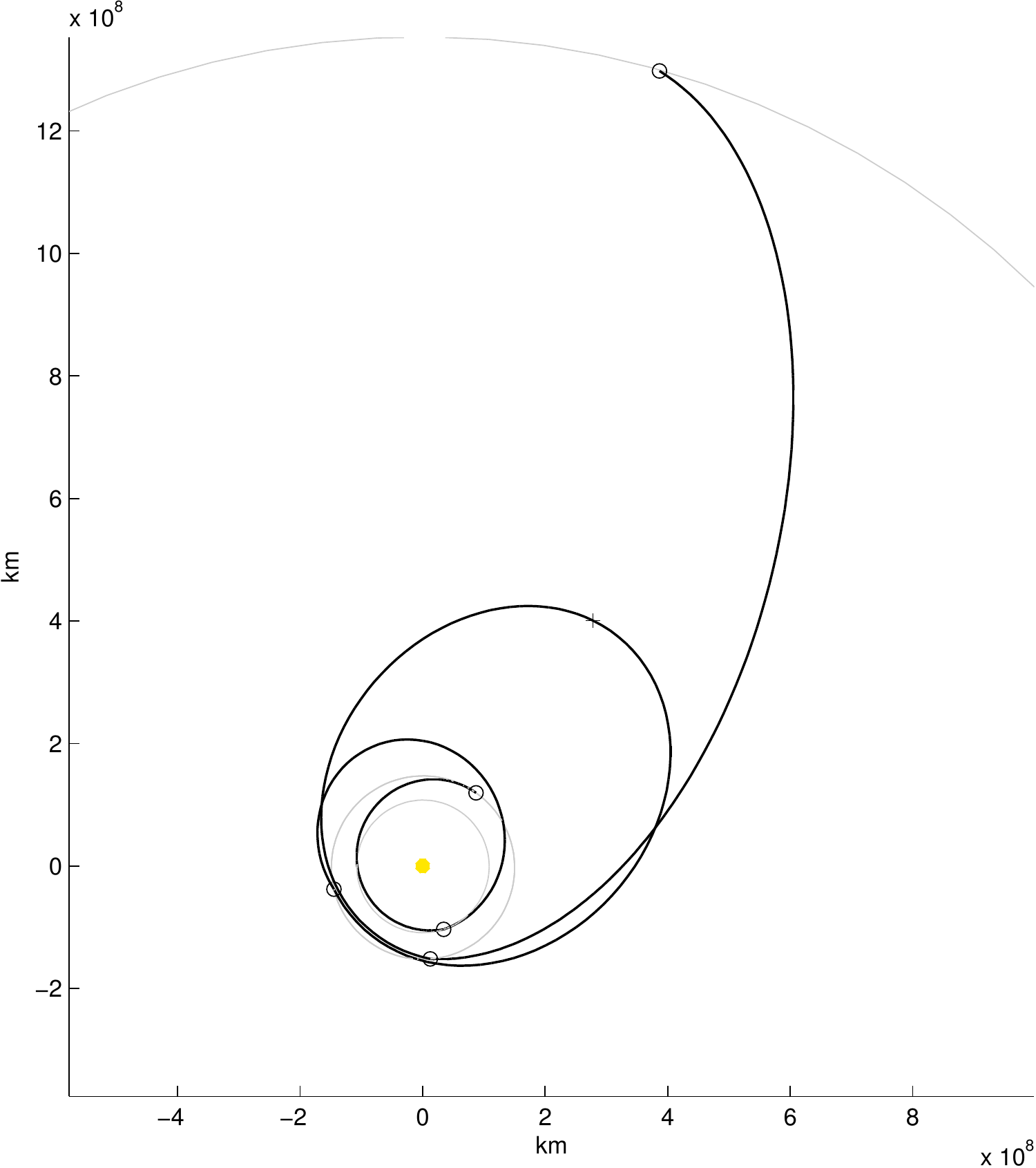}\label{fig:tandem}}

  \subfloat[][]{\includegraphics[width=10.9cm]{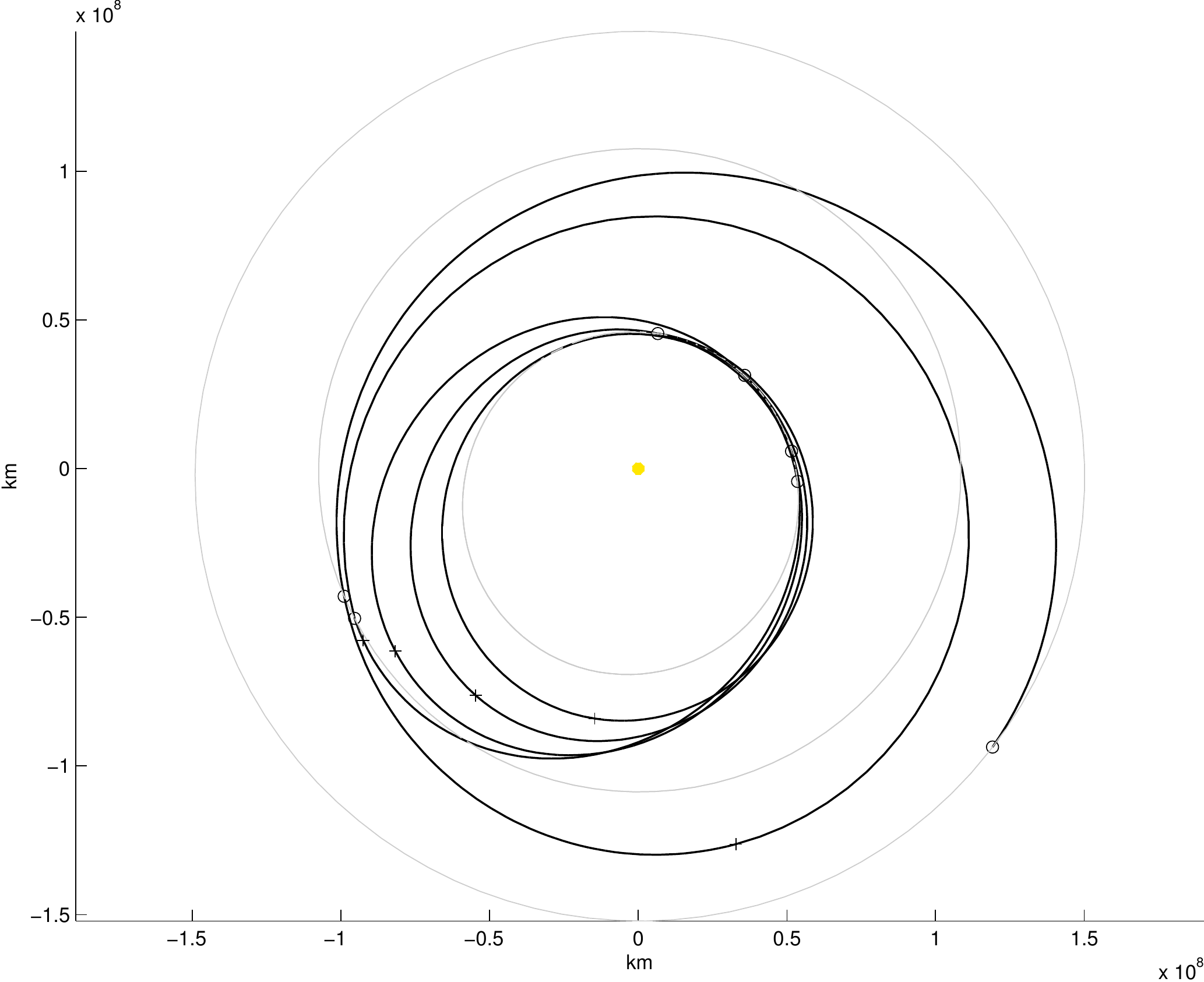}\label{fig:messengerfull}}
 \caption{Visualisation of the best trajectories found for the first three interplanetary trajectory problems: problem::cassini\_2 \subref{fig:cassini2}, problem::tandem(6,10) \subref{fig:tandem} and
problem::messenger\_full \subref{fig:messengerfull}.}
\end{figure*}

This problem represents the interplanetary trajectory of the spacecraft Cassini. For a detailed description of this global optimisation problem we refer the reader to the GTOP database \cite{gtop1,gtop2}. For
the purpose of this paper we just mention that the objective function represents the sum of all $\Delta V$, including the launch, where the last contribution (Jupiter arrival) is relative velocity with respect to Jupiter at arrival. For this problem we apply the fully automated search described above with three pruning cycles. At the end of the process (employing 7CPUs for a period of roughly 8 hours) the best trajectory found is visualized in Figure \ref{fig:cassini2}. Details on the trajectory as reported in Table \ref{tab:cassini2}. The trajectory is, essentially, the same best result posted in the database (22th May 2009)  and found by M. Schl{\"u}eter, J. Fiala, M. Gerdts at the University of Birmingham using the MIDACO solver developed within the project ``Non-linear mixed-integer-based Optimisation Technique for Space Applications'' co-funded by ESA Networking Partnership Initiative, Astrium Limited (Stevenage, UK) and the School of Mathematics, University of Birmingham, UK \cite{midaco}. The solution employs two deep space maneuvers during the first two legs as detailed in Table \ref{tab:cassini2}.

 \subsection{Results on problem::tandem(6,10)}

TandEM is one of the L-class candidate missions that were proposed in response to the European Space Agency call for proposals for the 2015-2025 Cosmic-Vision programme. Initially, the mission included a Titan orbiter and an Enceladus penetrator. The interplanetary part of the trajectory was preliminarly studied in 2008 by a Mission Analysis Outer Planets Working Group that included different experts from academia and space industry. In that preliminary study a baseline for the TandEM mission was defined and forms the basis of the problem here solved in an automated fashion using PaGMO. The baseline considers a launch with Atlas 501, and an injection orbit at Saturn with $e=0.985$, $r_p = 80330$ km. The mission objective is to maximize the final spacecraft mass at arrival and to complete the trajectory within ten years. For a detailed description of this global optimisation problem we refer the reader to the GTOP database \cite{gtop1,gtop2}. For this problem we apply the fully automated search described above with three pruning cycles. At the end of the process (employing 7CPUs for a period of roughly 6 hours) the best trajectory found is visualized in Figure \ref{fig:tandem}. Details on the trajectory are as reported in Table \ref{tab:tandem}. The solution found improves the previous best found by B. Addis, A. Cassioli, M. Locatelli, F. Schoen (from the Global Optimisation Laboratory, University of Florence) who also have the record on the solutions for all other TandEM problem instances (i.e. for different fly-by and time constraint). The final trajectory employs one only deep space manouvre during the Venus-Venus leg.

 \subsection{Results on problem::messenger\_full}

\begin{figure*}
 \begin{center}
  \includegraphics[width=14cm]{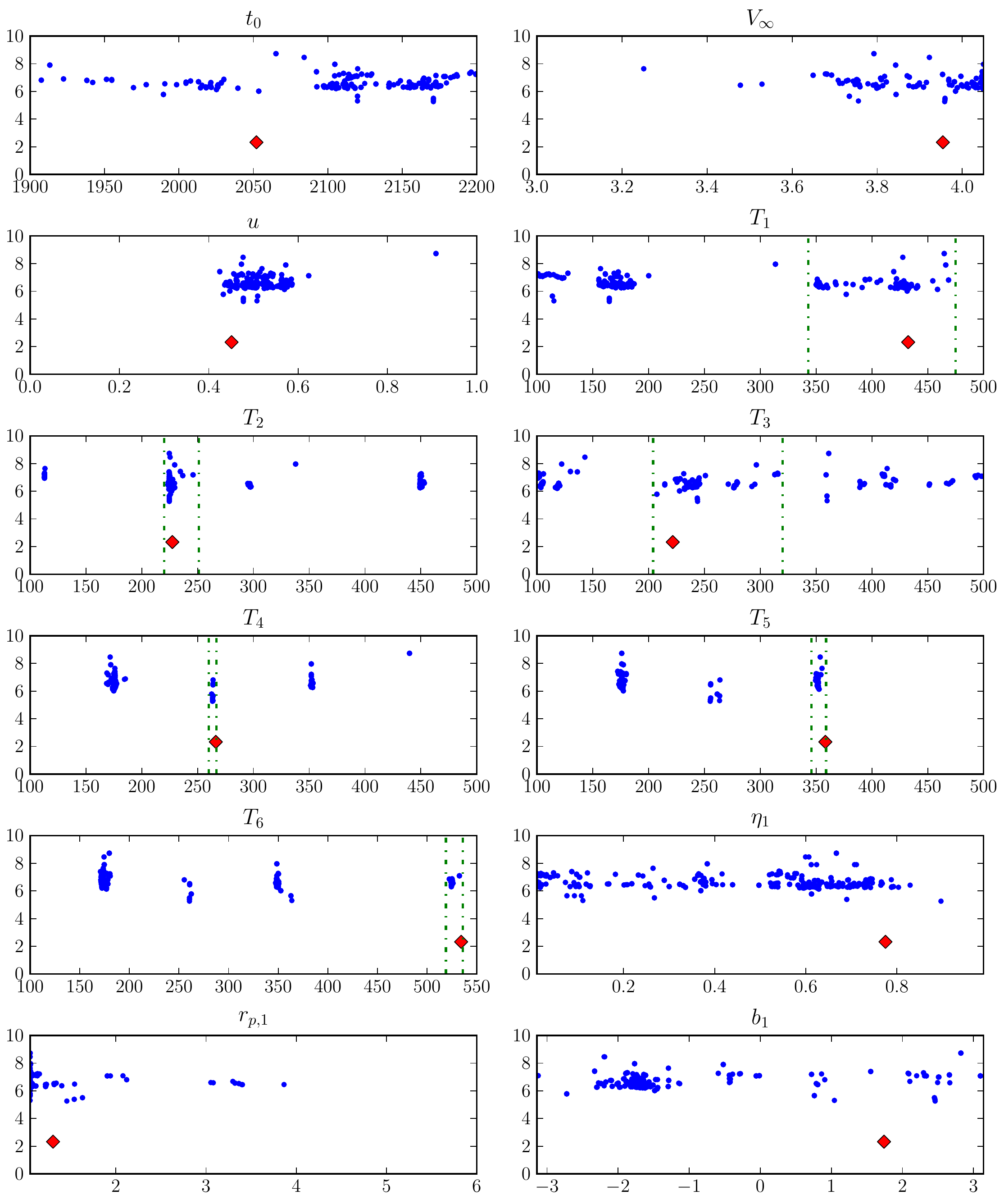}
  \caption{Manual pruning for a selection of variables from the Messenger problem. The blue dots represent the best candidate solutions of each run of the multistart strategy, the vertical green bars denote the new, narrowed bounds of the problem and the red diamond marker
represents the final solution. The clustering of the best candidate solutions in correspondence with the resonant flybys at Mercury is clearly visible for the variables $T_2$, $T_4$, $T_5$ and $T_6$.}
  \label{fig:pruning}
 \end{center}
\end{figure*}

This problem is probably the most complex problem in the GTOP database and represents one of the most complex chemical interplanetary trajectory ever designed and flown, that of the Messenger spacecraft. Messenger, at the time of writing, is on its way to Mercury, where (roughly next year, in 2011) will become the first spacecarft to ever orbit around the planet. Its path in the solar system to reach its final destination included a long planetary tour:
Earth-Earth-Venus-Venus-Mercury-Mercury-Mercury-Mercury. In the PaGMO version of the problem, the Messenger trajectory
is transcribed into a box-constrained global optimisation problem. The objective function is the total $\Delta V$ accumulated from the Earth launch to a Mercury Orbit Insertion (MOI) into an orbit having $e=0.704$, $r_p=2640$ km. For a detailed description of this global optimisation problem we refer the reader to the GTOP database \cite{gtop1,gtop2}. In this case, after a first run of the algorithm, cluster detection shows the presence of many different solution clusters related to the different possible resonances at Mercury, as shown in Figure \ref{fig:pruning}. The best solution after the first algorithm run is around $5.15$ km/sec. By focussing the optimisation into one of the detected clusters we obtain a solution at around $2.3$ km/sec which is lowering the GTOP database record substantially and is detailed in Table \ref{tab:messenger} and visualized in Figure \ref{fig:messengerfull}.

\subsection{Results on problem::sample\_return}

\begin{figure*}[ht]
 \centering
 \subfloat[][]{\label{fig:asteroids}\includegraphics[width=8cm]{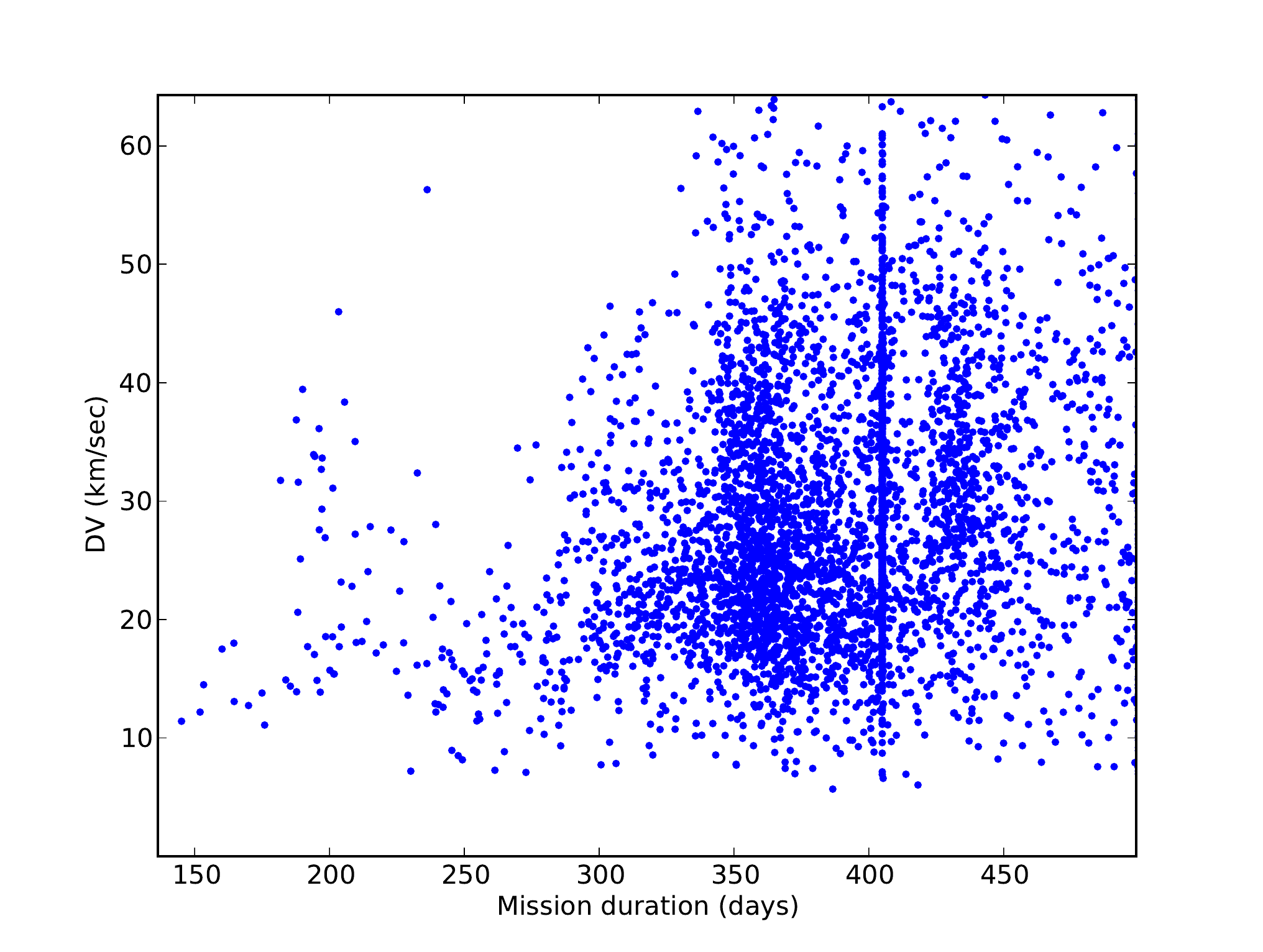}}
 \subfloat[][]{\label{fig:asteroids2}\includegraphics[width=8cm]{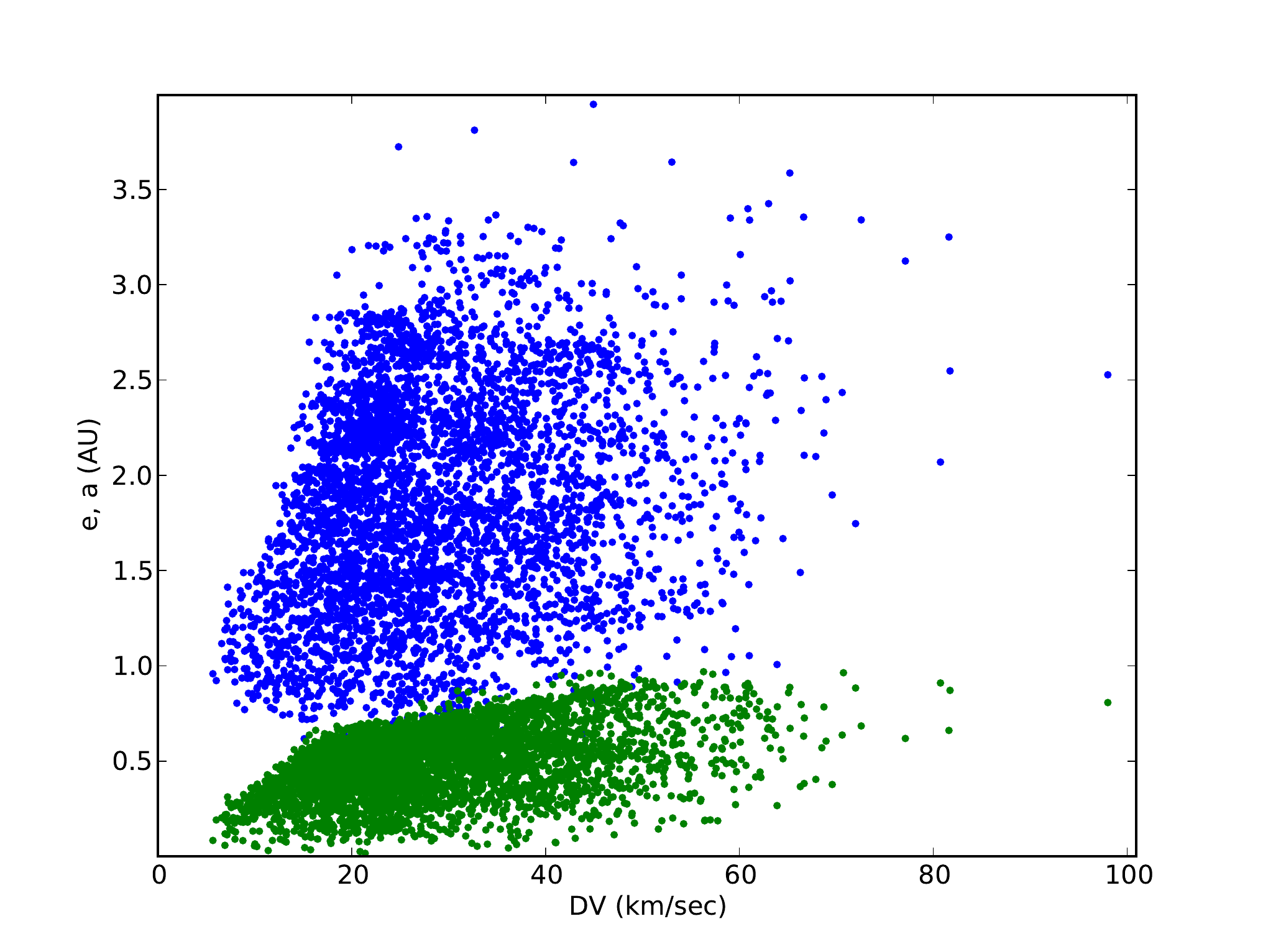}}
 \caption{Relations between $\Delta V$ and properties of the best solutions found: mission duration against $\Delta V$ \subref{fig:asteroids} and $\Delta V$ against semi-major axis and eccentricity \subref{fig:asteroids2}.}
 \label{fig:ast_recap}
\end{figure*}

A single instance of this problem represents an interplanetary trajectory starting from the Earth and performing a rendezvous with a selected asteroid. After a minimum waiting time the spacecraft is required to come back to the Earth with a maximum hyperbolic encounter velocity. One deep-space maneuver per leg was allowed, creating a global optimisation problem of dimension 12. This type of trajectory can be used to perform the preliminary selection of possible final asteroids for sample return missions. The same trajectory model is also relevant to design human missions to asteroids. For the purpose of this paper we do not enter into the details on the system design and launch window choice, instead it is our interest to `just pick an example' and show the possibility of using PaGMO to solve in a reasonable time a large number of problem instances (e.g. varying the final asteroid). We took all the asteroids listed in the JPL NEA database:\newline\newline
\url{http://neo.jpl.nasa.gov/cgi-bin/neo_elem}\newline\newline
having $H<22$, which roughly corresponds to asteroids having diameter larger than 200m. For the selected 4406 asteroids we optimised the trajectory considering a launch in the 2020-2050 time frame, allowing a final encounter velocity at the Earth return of 4.5. km/sec and minimizing the total $\Delta V$ as evaluated from:
\begin{multline*}
\Delta V = \Delta V_{L} + \Delta V_{dsm_1}+\Delta V_{R} + \\
\Delta V_{D} + \Delta V_{dsm_2} +\Delta V_{E},
\end{multline*}
where $ \Delta V_{L}$ is the hyperbolic velocity when leaving the Earth sphere of influence, $\Delta V_{dsm_1}$ is the first deep space maneuver, $\Delta V_{R}$ is the rendezvous velocity, $ \Delta V_{D}$ is the relative velocity at asteroid departure, $\Delta V_{dsm_2}$ is the second deep space maneuver and $\Delta V_{E}$ is the final braking maneuver to reduce the entry speed to 4.5 km/sec. A minimum waiting time on the asteroid of 5 days is also considered.

The computations where performed on an Xserve with 8 processing units and lasted 8 hours. The results are visualized in Figure \ref{fig:ast_recap}.

%% file: conclusions.tex
\section{Conclusions and future work}

In this paper we have presented PaGMO, a global optimisation software framework for parallel engineering optimisation developed within the Advanced Concepts
Team at the European Space Agency. We have tested PaGMO on three hard, realistic interplanetary trajectory optimisation problems (TandEM, Cassini and Messenger), showing how PaGMO is able to find automatically
the best known solutions for all three problems. With the help of a human-guided final pruning step, PaGMO was also able to locate the `real' trajectory for the Messenger
probe, consisting of multiple resonant flybys at Mercury. A fourth benchmark problem, consisting in the optimisation of simple trajectories to a selection of 4406 near-Earth asteriods,
was shown with the intent of highlighting PaGMO's parallelisation capabilities.

Future work on PaGMO will concentrate on areas such as:
\begin{itemize}
 \item extension of the computational capabilities via interfacing to popular massively parallel frameworks, such as MPI \cite{mpi} for scientific clusters
 and BOINC \cite{boinc} for distributed computing;
 \item exploration of the possibility of using GPGPU computing \cite{gpgpu} to speed-up the most time-consuming parts of the optimisation process;
 \item implementing/interfacing additional optimisation algorithms;
 \item interfacing with machine learning packages (such as PyBrain \cite{pybrain}), for easy coding of artificial intelligence problems.
\end{itemize}
Some of these activities will be tackled within the Google Summer of Code 2010, in which an international group of University students will be working during the summer on PaGMO under the mentorship of the
PaGMO development team - while being sponsored by Google.

PaGMO is Free Software, and it is available for download from the SourceForge website: \newline\newline
\url{http://pagmo.sourceforge.net}